\documentclass[fleqn,usenatbib]{mnras}

\usepackage{newtxtext,newtxmath}

\usepackage[T1]{fontenc}

\DeclareRobustCommand{\VAN}[3]{#2}
\let\VANthebibliography\thebibliography
\def\thebibliography{\DeclareRobustCommand{\VAN}[3]{##3}\VANthebibliography}

\usepackage[dvipsnames]{xcolor}

\usepackage{graphicx}	
\usepackage{amsmath}	
\usepackage{hyperref}
\usepackage{threeparttable}

\newcommand{\GWT}{\texttt{GW-Universe Toolbox}}

\title[BNS merger rate density history]{A systematic study of binary neutron star merger rate density history using simulated gravitational wave and short gamma-ray burst observations}

\author[Y. F. Du et al.]{
Y. F. Du,$^{1,2}$
E. S. Yorgancioglu,$^{1,2}$
S. X. Yi,$^{1}$\thanks{E-mail: sxyi@ihep.ac.cn }
T. Y. Cao,$^{1,2}$
S. N. Zhang,$^{1,2,3}$
\\
$^{1}$Key Laboratory of Particle Astrophysics, Institute of High Energy Physics, Chinese Academy of Sciences, 19B Yuquan Road, Beijing 100049, People’s Republic of China\\
$^{2}$University of Chinese Academy of Sciences, Chinese Academy of Sciences, Beijing 100049, People’s Republic of China\\
$^{3}$National Astronomical Observatories, Chinese Academy of Sciences, Beijing 100012, People’s Republic of China
}

\date{Accepted XXX. Received YYY; in original form ZZZ}

\pubyear{\the\year{}}

\begin{document}
\label{firstpage}
\pagerange{\pageref{firstpage}--\pageref{lastpage}}
\maketitle

\begin{abstract}
Measuring the merger rate density history of binary neutron stars (BNS) can greatly aid in understanding the history of heavy element formation in the Universe. Currently, second-generation Gravitational Wave (GW) detectors can only measure the BNS merger rate density history at low redshifts ($z$ $\sim$ 0.1). Short gamma-ray bursts (sGRBs) may trace the BNS merger to higher redshifts ($z$ $\sim$ 3). However, not all BNS mergers result in sGRBs, and it is not certain that all sGRBs originate from BNS mergers. In this study, we simultaneously utilize simulated BNS merger GW signals detected by the advanced LIGO design and sGRB signals detected by {\it Fermi}/GBM to constrain the BNS merger rate density history up to $z$ $\sim$ 3.
The results indicate that with $\sim$ 8 GWs and 571 sGRBs, the BNS merger rate density can be measured with an accuracy of about 50\% through $z=0$ to $z=1$. The ratio of the jet opening angle-corrected sGRB event rate density to the BNS merger rate density, denoted as $\eta$, can be constrained to a relative uncertainty of 45\%. With $\sim$ 21 GWs and 761 sGRBs, the BNS merger rate density can be measured to approximately 35\% and 40\% at $z=0$ and $z=1$, respectively. Meanwhile, $\eta$ can be constrained to a relative uncertainty of 28\%. Additionally, in our parameterized simulation, we find that at least approximately $\sim$550 sGRBs are needed to constrain the characteristic delay time in the star formation rate model, given a relative error of 50\% in the estimated redshift.
\end{abstract}




\section{Introduction} \label{sec:intro}

Gravitational Waves (GWs) and electromagnetic waves emanating from binary neutron star (BNS) mergers offer a unique opportunity to probe the complex physical processes at play in these cosmic events, significantly broadening our astrophysical comprehension \citep[see][for a detailed review]{2017ApJ...848L..12A, 2021ARA&A..59..155M}. In particular, the association of GRB 170817A with GW170817 verifies
that at least a fraction of, if not all, short gamma-ray bursts (sGRBs) originate from the mergers of BNSs \citep[]{2014ARA&A..52...43B, 2021NatAs}. The kilonova AT2017 gfo, which is associated with GW170817, confirms the occurrence of the r-process following the BNS merger, a crucial astrophysical process in the Universe that can generate nuclei of elements heavier than the iron peak \citep[]{1982ApL....22..143S, 2024MNRAS.529.1154C}. 
Understanding the BNS merger rate is thus critical for predicting the frequency of such events and for shedding light on the synthesis of heavy elements in the Universe. 

BNS mergers can give rise to various astronomical phenomena that we can observe, including GWs, sGRBs, and kilonovae. Thus, these astronomical phenomena can be used to infer the BNS merger rates. Currently, some studies have provided constraints on BNS merger rates based on observational data from GWs, sGRBs, and kilonovae candidates. The BNS merger rate density, obtained by constraining the BNS population through the second Gravitational-Wave Transient Catalog (GWTC-2) and third Gravitational-Wave Transient Catalog (GWTC-3) released by the LIGO/Virgo Collaboration, are $320_{-240}^{+490}\,\rm{Gpc^{-3}\,yr^{-1}}$\citep[]{2021ApJ...913L...7A} and $10-1700\, \rm{Gpc^{-3}\,yr^{-1}}$\citep[]{2023PhRvX..13a1048A}, respectively. Due to the detection range of GWs from BNS mergers of LIGO/Virgo ($\sim$ 100 Mpc), the aforementioned merger rate density measured through GW catalogs are considered to be local. \cite{2012MNRAS.425.2668C} obtained an sGRB rate density of $8_{-3}^{+5}-1100_{-470}^{+700}\,\rm{Gpc^{-3}\,yr^{-1}}$ based on a sample of 8 sGRBs observed by Swift with well-determined redshift. The application of their method is greatly challenged by the complex instrument selection effects and sGRB beaming constraints. \cite{2013ApJ...767..140P} also estimated the NS-NS/NS-black hole (BH) merger rate density. They obtained the coalescence rates in the local universe ranging from 500 to 1500 $\rm{Gpc^{-3}\,yr^{-1}}$. However, their result is strongly dependent on the weakly constrained jet opening angle of the collimated emission from the sGRBs. \cite{2015ApJ...815..102F} and \cite{2023ApJ...959...13R} used the afterglows of sGRBs to constrain the jet opening angle, and their results yielded a local sGRB event rate density of $270_{-180}^{+1580}\,\rm{Gpc^{-3}\,yr^{-1}}$ and $360-1800\, \rm{Gpc^{-3}\,yr^{-1}}$. These constraints pertain to average local merger rates rather than merger rate density as function of redshift.

In addition to the aforementioned methods, the stochastic gravitational wave background (SGWB) can also be employed to constrain the BNS merger rate density history. Due to instrumental sensitivity limitations, second-generation ground-based GW detectors typically observe numerous BNS merger signals with low signal-to-noise ratio (S/N), which become indistinguishable from the background noise and collectively contribute to the SGWB. These unresolved BNS merger signals produce characteristic enhancements in the SGWB energy density spectrum within specific frequency bands \citep{2018PhRvL.120i1101A, 2022PhRvD.105l3522A}. The unique spectral imprints from BNS mergers in the SGWB energy spectrum enable their astrophysical identification. The cosmic star formation history and the delay time distribution (DTD) between BNS system formation and merger significantly influence the amplitude of BNS merger signals in the SGWB. Consequently, SGWB measurements can be utilized to constrain the parameter space of DTD models. \cite{2020ApJ...901..137S} demonstrated that DTD variations induce five orders-of-magnitude differences in BNS merger signal amplitudes within the SGWB energy spectrum, though such signals remain undetectable by aLIGO's sensitivity. Third-generation ground-based  detectors like Einstein Telescope (ET), and space-based observatories like Laser Interferometer Space Antenna (LISA) are projected to detect these BNS merger signatures in the SGWB, thereby constraining the BNS merger history \citep{2019ApJ...871...97C, 2019ApJ...878L..13S}.

Since the redshift of detected sGRBs reaches up to 3, the constraints on the BNS merger rate density at redshifts up to 3 can be derived from these detected sGRBs \citep{2023ApJ...949L...4L}. 
The GW observations primarily provide constraints on the BNS merger rate density at redshift below 0.1. There is also a scarcity of samples with redshift below 0.1 for sGRBs. Therefore, combining these two methods with GW and sGRB data allows for a better understanding of the evolution of the BNS merger rate density across redshift ranging from around 0 to 3. However, the primary issue is that the sGRB rate derived from these sGRBs does not directly reflect the BNS merger rate. This discrepancy arises due to the following reasons: (1) The jet opening angles of sGRBs are difficult to determine, which significantly impacts the estimation of the sGRB rate; (2) only a subset of sGRBs may actually originate from BNS mergers and BNS mergers do not necessarily produce sGRBs. In our model, we have assumed the distribution of the jet opening angles for sGRBs, ultimately deriving the jet opening angle-corrected sGRB event rate density. Consequently, the first reason does not need to be considered in our simulations.
For the second reason, the joint data of GW and sGRBs can constrain  the ratio of the jet opening angle-corrected sGRB event rate density to the BNS merger rate density. 


To evaluate the feasibility of this method, this study employs simulated catalogs of GW and sGRB sources to constrain the BNS merger rate density. We generate synthetic GW catalogs with varying S/N thresholds and detectors using the software package \GWT\, \citep[][]{2022A&A...663A.155Y}, and sGRB catalogs using the methodology described in \cite{2024MNRAS.534.2715D}. Bayesian inference is then applied to these simulated catalogs to estimate the BNS merger rate density. We assume that the likelihood functions of the GW and sGRB observations can be combined multiplicatively to represent the total likelihood, since GW and sGRB observations are independent.

In Section 2, we will introduce the BNS population and sGRB model, along with the process for generating the GW and sGRB catalogs. Section 3 will detail the Bayesian method used to constrain the BNS merger rate density from these catalogs. Section 4 will present the results obtained through Bayesian inference. Finally, Section 5 will give a discussion of the findings and outlook for future research.

\section{Simulated events} \label{sec:simu}

\subsection{BNS population and sGRB model}

\begin{figure}
    \centering
    \includegraphics[width=.48\textwidth]{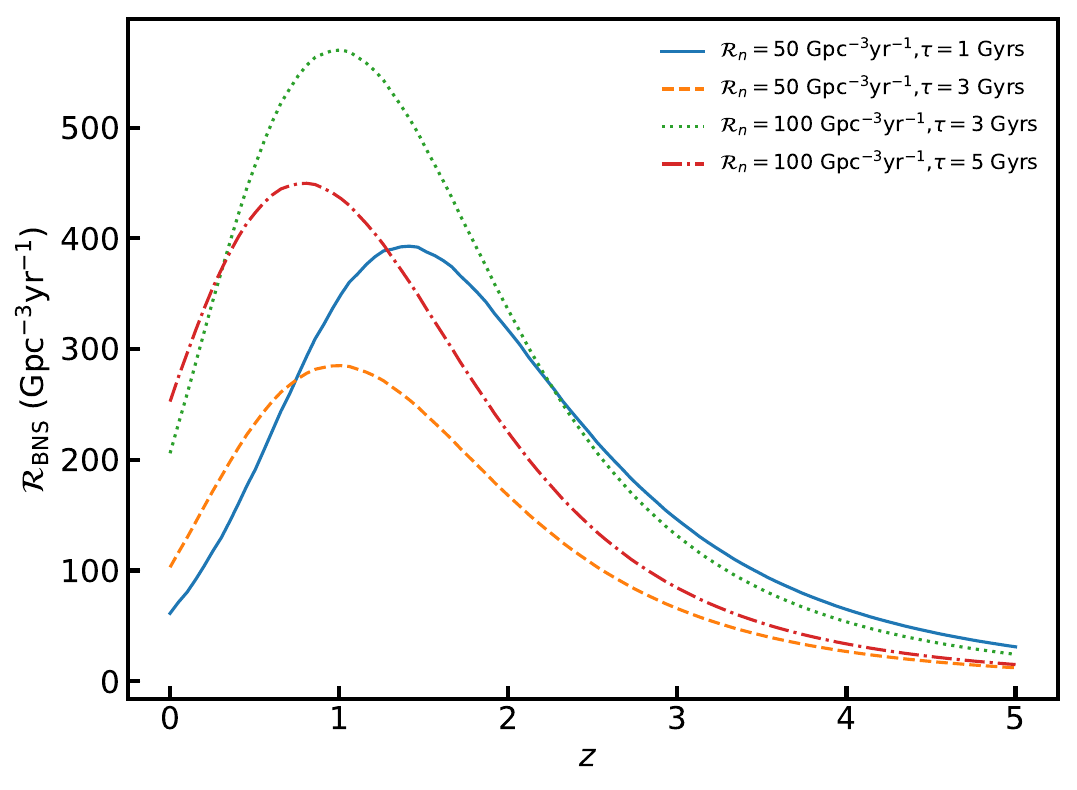}
    \caption{The BNS merger rate density as the function of redshift under different $\mathcal{R}_{n}$ and $\tau$.}
    \label{fig:bns_rate}
\end{figure}

The differential BNS merger rate as function of redshift $R(z)\equiv \frac{dN}{dtdz}$, can be written in terms of the volumetric total BNS merger rate density $\mathcal{R}(z)\equiv \frac{dN}{dtdV_{c}}$ in the source frame as 

\begin{equation}
R(z)=\frac{1}{1+z}\frac{dV_{c}}{dz}\mathcal{R}(z),
\label{r_r}
\end{equation}
where $dV_{c}/dz$ is the differential comoving volume, and the $(1+z)^{-1}$ term arises from converting source-frame time to detector-frame time. $\mathcal{R}(z)$ is often referred to as the cosmic BNS merger rate density:
\begin{equation}
\mathcal{R}(z_{\rm m})=\mathcal{R}_n\int_{z_{\rm m}}^{\infty}\psi(z_{\rm f})P(z_{\rm m}|z_{\rm f})dz_{\rm f},
\label{merger_rate}
\end{equation}
where $\psi(z_{\rm f})$ is the non-normalized Madau-Dickinson star formation rate:
\begin{equation}
\psi(z)=\frac{(1+z)^{\alpha}}{1+(\frac{1+z}{C})^{\beta}},
\label{Madau-Dickinson}
\end{equation}
with $\alpha=2.7$, $\beta=5.6$, $C=2.9$ \citep{2014ARA&A..52..415M}, and $P(z_{\rm m}|z_{\rm f})$ is the probability that a BNS merges at $z_{\rm m}$  given that the binary was formed at $z_{\rm f}$.  This is referred to as the distribution of delay times, which takes the form: 
\begin{equation}
P(z_{\rm m}|z_{\rm f},\tau)=\frac{1}{\tau}{\rm exp}[-\frac{t_{\rm f}(z_{\rm f})-t_{\rm m}(z_{\rm m})}{\tau}]\frac{dt}{dz}.
\label{prob_merger}
\end{equation}
In the above equation, $t_{\rm f}$ and $t_{\rm m}$ are the look-back times as functions of $z_{\rm f}$ and $z_{\rm m}$, respectively. $\tau$ is the characteristic delay time. In this study, $\mathcal{R}_{n}$ and $\tau$ are the pair of hyperparameters that define the merger rate density of the population. Figure \ref{fig:bns_rate} illustrates several curves depicting the merger rate density as it varies with redshift under different $\mathcal{R}_n$ and $\tau$ conditions. From this figure, it is evident that as $\mathcal{R}_n$ increases, the merger rate density also rises. Additionally, as $\tau$ increases, the curvature of the rate density curves becomes more pronounced.

In addition to the redshift dependence, the BNS merger rate density also depends on the mass distribution of the neutron stars in the binary system. Therefore, the BNS merger rate density can be expressed as:
\begin{equation}
\mathcal{R}_{\rm BNS}(z, m_1, m_2) = p(m_1) p(m_2) \mathcal{R}(z)
\end{equation}
where $p(m_1)$ and $p(m_2)$ are the probability density functions for the masses $m_1$ and $m_2$, respectively. The combined redshift and mass dependence is included in the formulation of $\mathcal{R}_{\rm BNS}(z, m_1, m_2)$, which defines the full population model for BNS mergers. We use the \GWT\, to generate GW catalogues. We select the hyperparameters for the BNS merger rate density as $\mathcal{R}_{n}=50\,\rm Gpc^{-3}\,yr^{-1}$, consistent with \cite{2023A&A...672A..74H}, and $\tau=3\,\rm Gyrs$. For the distribution of Neutron Star (NS) masses, we use truncated Gaussian distributions with $m_{\rm{mean}}=1.4\,M_\odot$, $m_{\rm{scale}}=0.5\,M_\odot$, $m_{\rm{low}}=1.1\,M_\odot$, $m_{\rm{high}}=2.5\,M_\odot$, consistent with \cite{2022A&A...663A.155Y}.  
The GW detector we choose is the advanced LIGO (aLIGO) design.

By applying the BNS population model described above using the methodology in \cite{2024MNRAS.534.2715D}, we can obtain the entire BNS merger population within a certain redshift range. The sGRB model we choose is consistent with \cite{2024MNRAS.534.2715D}, corresponding to a structured Gaussian-type jet model. Then the luminosity per solid angle in the observer frame is
\begin{equation}
    L(\theta_c,\theta_v)=\frac{dE}{dtd\Omega}=\frac{L_0}{4\pi}e^{-\theta_v^2/\theta_c^2},
    \label{eq:Lum}
\end{equation}
where $\theta_c$ is the jet’s core angle, $L_0$ is the isotropic-equivalent luminosity observed on-axis, and $\theta_v$ is viewing angle. The flux for a GRB at redshift $z$ and viewing angle $\theta_v$ is:
\begin{equation}
F(\theta_c,\theta_v,z)=\frac{L(\theta_c,\theta_v)}{4\pi D^2_{L}(z)},
\end{equation}
where $D_{L}(z)$ is the luminosity distance converted from $z$ by a cosmological model found in \cite{2020A&A...641A...1P}. Now we assume that the detection flux threshold is $F_{\rm lim}$. For \textit{Fermi} Gamma-ray Burst Monitor (GBM),  we take its 64 ms limiting flux, $F_{\rm lim}=2.0\times 10^{-7}$\,erg\,cm $^{-2}$\,s$^{-1}$ \citep{2017ApJ...848L..14G}. Then the detectable minimum isotropic luminosity is
\begin{equation}
L_{\rm min}=F_{\rm lim}4\pi D_{L}^2(z),
\end{equation}
that means $L_0$ should satisfy
\begin{equation}
\textcolor{black}{L_0 \ge L_{\rm lim}
  \;\equiv\;
  F_{\rm lim}\,(4\pi)^2\,D_{L}^2(z)\,\exp(\theta_v^2/\theta_c^2)\,.}
\end{equation}

where $L_{\rm lim}$ is the minimum on-axis luminosity as defined in equation \ref{eq:Lum}. The $L_0$ distribution of the structured jet in the sGRB has been investigated in a study by \cite{2023A&A...680A..45S}. Based on their work, we adopt a power law with an index of $-A$, with a lower exponential cutoff below $L_{\star}$ in the distribution of $L_0$ as follows
\begin{equation}
\Phi(L_0)=\frac{\Phi_0 A}{L_{\star}(1-1/A)}\exp\left[-\left(\frac{L_{\star}}{L_0}\right)^A \right]\left(\frac{L_0}{L_{\star}}\right)^{-A},
\label{dis_L}
\end{equation}
where $\Phi_0$ is the normalization parameter. Following  \cite{2023A&A...680A..45S}, we take the values $A=3$ and $L_{\star}=0.5\times 10^{52}$\,erg\,s$^{-1}$.
Then the detection probability is:
\begin{equation}
\label{eq:p_det1}
P_{\rm{det, sGRB}}(\theta_c,\theta_v,z)\propto \frac{\int_{L_{\rm{lim}}}^\infty\Phi(L_0)dL_0}{\int_{0}^\infty\Phi(L_0)dL_0}.
\end{equation}
In equation \ref{eq:p_det1}, the results of the two definite integrals are the gamma function $\Gamma$ and the incomplete gamma function $\widetilde{\Gamma}$, respectively. Thus, we get
\begin{equation}
P_{\rm{det, sGRB}}(\theta_c,\theta_v, z)=\Lambda \frac{\widetilde{\Gamma}(1-1/A, (L_{\star}/L_{\rm lim})^A)}{\Gamma(1-1/A)}.
\label{eq:Pdet_grb}
\end{equation}
where $\Lambda=0.6$, which means a 60 percent sky coverage (time-averaged) for \textit{Fermi} GBM telescope, following \cite{2016ApJ...818..110B}. We assume that $\theta_c$ has a Gaussian distribution with a mean of 1.5$^{\circ}$ and the a standard deviation of 2$^{\circ}$, which is in accordance with the range from 1$^{\circ}$ to 9$^{\circ}$ implied from  observations \citep{2018Natur.561..355M}. The resultant {\it Fermi} GBM detection rate and redshift range are in good agreement with reality (39 sGRB/year and up to $z\sim3$).

\begin{figure}
    \centering
    \includegraphics[width=.48\textwidth]{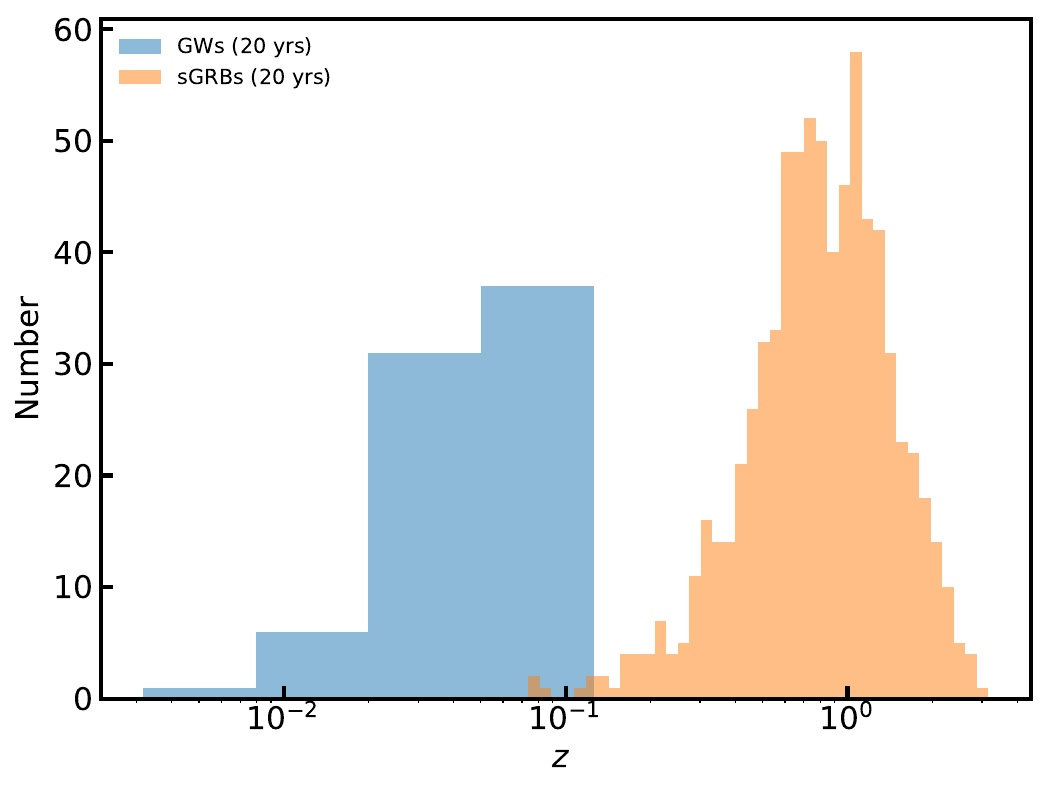}
    \caption{The redshift histogram of the simulated GWs and sGRBs. The observation durations for both GWs and sGRBs are 20 years. The GW detector is aLIGO design and S/N threshold is 8. The GRB detector is Fermi/GBM. The total number of GWs and sGRBs are 75 and 761.}
    \label{fig:hist}
\end{figure}

We plot the redshift distribution for GWs and sGRBs utilized in our simulation in Figure \ref{fig:hist}. The redshift of the GWs is predominantly below 0.1, while the redshift of sGRBs ranges from 0 to 3. Over a 20-year observation period by {\it Fermi}/GBM, the number of sGRBs amounts to 761, aligning well with the actual rate of 39 sGRBs per year.

\subsection{The redshift of the GW and GRB events}

In order to represent the uncertainties of the redshift, we assume that the measured redshifts are randomly selected from a lognormal distribution with the simulated redshifts and their errors as the mean and variance, respectively. The redshift of simualted GWs is determined by their luminosity distance, which enables the estimation of redshift error based on the error in the luminosity distance. We employ the approximation $\Delta D \sim D/\rho$ to estimate the error in $D_{\rm L}$ \citep[]{1994PhRvD..49.2658C}, where $\rho$ represents the S/N of the GW. Consequently, in a low-redshift approximation, the redshift error of GWs, denoted as $\Delta z_{\rm GW}$, is approximately $z_{\rm GW}/\rho$. 
Conversely, the redshift error of a sGRB is typically determined via spectroscopic observations of the host galaxy, leading to minimal errors that can be disregarded. Nevertheless, a significant portion of sGRBs, approximately 90\%, lack redshift measurements.  
We assume that the redshift distribution of sGRBs without redshift measurements is consistent with that of sGRBs with redshift measurements, which will be discussed in Section 4. Many studies have used the isotropic equivalent energy, duration, peak energy, and other parameters of GRBs to predict their redshifts \citep[]{2018Ap&SS.363..223Z, 2023ApJ...943..126D, 2024MNRAS.529.2676A}. We adopt the method proposed in the \cite{2024ApJS..271...22D}, where the estimated redshift errors are approximately half of the redshift values. Therefore, in our simulations, we randomly choose 90\% of the simulated GRBs and set the relative error of their redshifts to be 50\%. The remaining GRBs are assumed to have accurate redshift measurements (with zero uncertainties). In our study, we find that the magnitude of redshift errors can impact the constraints on the BNS merger rate history, which will be further elaborated in Section 4.

\section{Methods} \label{sec:method}
We assume that BNS mergers follow an inhomogeneous Poisson process with rate density $\mathcal{R}_{\rm GW}(m_1,m_2,z| \vec{\lambda})$ depending on some hyperparameters $\vec{\lambda}$, which can be written as

\begin{equation}
\mathcal{R}_{\rm GW}(m_1,m_2,z | \vec{\lambda})=p(m_1)p(m_2)\mathcal{R}_{\rm GW}(z|\vec{\lambda}),
\label{R_gw}
\end{equation}
where $p(m_{1,2})$ is the mass function of the neutron star in the binary, which is assumed to be a truncated Gaussian distribution with upper and lower cuts.  
The detection rate of the BNS merger GW can be written as
\begin{equation}
\begin{split}
\dot{N}_{\rm det, GW}(m_1, m_2,z | \vec{\lambda})=\alpha_{\rm GW}(m_1, m_2,z)
\mathcal{R}_{\rm GW}(m_1, m_2, z | \vec{\lambda})\frac{1}{1+z}\frac{dV_{c}}{dz},
\label{NGW_det}
\end{split}
\end{equation}
where $\alpha_{\rm GW}$ is the fraction of GW sources that are detectable by our experiment, according to some detection threshold. For GW, $\alpha_{\rm GW}$ can be calculated using the equation (19) in \cite{2022A&A...663A.155Y}.

From simulation, we get the GW data $d_{\rm GW} \equiv\left\{z_i, m_{1, i}, m_{2, i}\right\}_{i=1}^{N}$, where $N$ is the number of the events. The occurrence of these events follows an inhomogeneous Poisson process, similar to that of sGRBs. For a single event in the dataset, the probability density that it will occur is given by:

$$\frac{\dot{N}_{\rm det, GW}(m_{1, i}, m_{2, i},z_i | \vec{\lambda}) \Delta T}{\Lambda _{\rm GW}},$$
where 
\begin{equation}
\Lambda_{\rm GW} \equiv \int \dot{N}_{\rm det, GW}(m_1, m_2,z | \vec{\lambda})dm_1dm_2dz\Delta T,
\label{int_GW}
\end{equation}
which is the expected number of detections. $\Delta T$ is the duration time of the observation. For all events in the data, we multiply them all together, i.e., $$\prod^N_{i=1}\frac{\dot{N}_{\rm det, GW}(m_{1, i}, m_{2, i},z_i | \vec{\lambda}) \Delta T}{\Lambda _{\rm GW}}.$$ Moreover, since these detection events will occur in a certain order, the term should be multiplied by $N!$. Therefore, the probability of all events in the $d_{\rm GW}$ occurring is 
\begin{equation}
P_{d_{\rm GW}}=\left[\prod^N_{i=1}\frac{\dot{N}_{\rm det, GW}(m_{1, i}, m_{2, i},z_i | \vec{\lambda}) \Delta T}{\Lambda _{\rm GW}}\right]N!.
\label{inhomo1}
\end{equation}

In a Poisson process with expected event number $\Lambda_{\rm GW}$, the probability of observing an event number $N$ is: 
\begin{equation}
P_N=\frac{\Lambda_{\rm GW}^N}{N!}e^{-\Lambda_{\rm GW}}.
\label{inhomo2}
\end{equation}
By multiply equation \ref{inhomo1} and equation \ref{inhomo2}, we obtain the likelihood of the data given the population-level hyperparameters $\vec{\lambda}$ and $m_1, m_2$: 
\begin{equation}
\begin{split}
\mathcal{L}_{\rm GW}(d_{\rm GW} | \mathcal{R}_{\rm GW})= \left [\prod^N_{i=1}\dot{N}_{\rm det, GW}(m_{1, i}, m_{2, i},z_{i} | \vec{\lambda})\Delta T\right] e^{-\Lambda_{\rm GW}}.
\end{split}
\label{like_gw}
\end{equation}

Similarly, we assume that the production of BNS merger sGRB sources is an inhomogeneous Poisson process with rate density $\mathcal{R}_{\rm sGRB}(\theta_{c}, z | \vec{\lambda})$ depending on some hyperparameters $\vec{\lambda}$, which also can be written as
\begin{equation}
\mathcal{R}_{\rm sGRB}(\theta_{c}, z | \vec{\lambda})=f(\theta_{c})\mathcal{R}_{\rm sGRB}(z|\vec{\lambda}),
\label{r_sGRB}
\end{equation}
where $f(\theta_{c})$ is the distribution of the jet’s core angle.  

The detection rate of sGRB can be written as 
\begin{equation}
\begin{split}
    \dot{N}_{\rm det, sGRB}(\theta_{c}, z | \vec{\lambda})= \alpha_{\rm sGRB}(\theta_{c}, z) 
    \mathcal{R}_{\rm sGRB}(z| \vec{\lambda})f(\theta_{c})\frac{1}{1+z}\frac{dV_{c}}{dz},
\end{split}
\label{NGRB_det}
\end{equation}
where $\alpha_{\rm sGRB}(\mathcal{\theta}_{c}, z)$ is the fraction of sGRB sources that are detectable by the GRB detector employed, based on equation \ref{eq:Pdet_grb}, $$\alpha_{\rm{sGRB}}(\theta_c, z)=\int P_{\rm{det, sGRB}}(\theta_c,\theta_v,z)\sin{\theta_v}d\theta_v.$$   

From simulation, we get the sGRB data $d_{\rm sGRB} \equiv\left\{z_i, \theta_{c, i} \right\}_{j=1}^{M}$, where $M$ is the number of the events.   
For the inhomogeneous Poisson process, the likelihood of the data given the population-level hyperparameters $\vec{\lambda}$, and $\theta_{c}$ is 
\begin{equation}
\begin{split}
  \mathcal{L}_{\rm sGRB}(d_{\rm sGRB} | \mathcal{R}_{\rm sGRB})= \left [\prod^M_{j=1}\dot{N}_{\rm det, sGRB}(\theta_{c,j},z_{j} | \vec{\lambda})\Delta T\right] e^{-\Lambda_{\rm sGRB}},  
\end{split}
\label{like_sGRB}
\end{equation}
where $$\Lambda_{\rm sGRB} \equiv \int \dot{N}_{\rm det, sGRB}(\theta_{c}, z  | \vec{\lambda})d\theta_{c}dz\Delta T,$$ $\Delta T$ is the duration time of the observation.

The relationship among BNS merger rate density $\mathcal{R}_{\rm BNS}(z|\vec{\lambda})$, intrinsic BNS merger GW rate density $\mathcal{R}_{\rm GW}(z|\vec{\lambda})$ and intrinsic BNS merger sGRB rate density $\mathcal{R}_{\rm sGRB}(z|\vec{\lambda})$ is
\begin{equation}
\mathcal{R}_{\rm GW}(z|\vec{\lambda})=\mathcal{R}_{\rm BNS}(z|\vec{\lambda})=\mathcal{R}_{\rm sGRB}(z|\vec{\lambda})/\eta,
\label{prior1}
\end{equation}
where $\eta$ is the ratio of the jet opening angle-corrected sGRB event rate density to the BNS merger rate density. However, the observed value of $\eta$ can be either larger or less than unity for two reasons: 
(1) Beside originated from BNS mergers, sGRBs may also originate from BH-NS mergers, or non-compact star mergers \citep{2021NatAs...5..911Z}, such as massive stellar collapses or magnetar flares. They could also be long gamma-ray bursts (lGRBs) that appear short due to the tip-of-the-iceberg effect. These scenarios would lead to $\eta > 1$. 
(2) BNS mergers do not necessarily produce sGRBs. For instance, weak magnetic fields may prevent the formation of ultra-relativistic jets, the merger may directly form a black hole, or the jet may fail to break out. These cases would result in $\eta < 1$. 
However, studies on $\eta$ are currently limited, and it may also be a function that varies with redshift. The joint likelihood of all GW and sGRB data is 
\begin{equation}
\mathcal{L}_{\rm GW, sGRB}(d_{\rm GW},d_{\rm sGRB}| \mathcal{R}_{\rm BNS}(z|\vec{\lambda}), \eta)=\mathcal{L}_{\rm GW}\times\mathcal{L}_{\rm sGRB}.
\label{likelihood_BNS}
\end{equation}
In our above treatments, there are assumptions that the neutron star mass function, sGRB jet's core angle, and $\eta$ do not depend on redshift. Therefore, $p(m_1)$, $p(m_2)$ and $f(\theta_{c})$ have been separated from the merger rate density, resulting in the merger rate density being solely a function of redshift. The redshift errors are accounted for in the likelihood function in the following way: for each source in the catalogue, we resample its redshift according to a probability distribution defined by its mean value and uncertainty. We compute a likelihood for each realization, and we average the likelihood over all realizations.

\section{Results}

\begin{table}
	\centering
	\caption{Prior boundaries and injected values of these parameters}
	\label{tab:2_table}
	\begin{threeparttable}[b]
	\begin{tabular}{cccc} 
		\hline
		& $\mathcal{R}_{n}$ (${\rm Gpc^{-3}\,yr^{-1}}$) & $\tau$ (${\rm Gyr}$) & $\eta$ \\
		\hline
		Prior & 1-1000 & 0.12-10 & 0.01-10 \\
		\hline
		Injected & 50 & 3 & 1 \\
		\hline
	\end{tabular}
    \end{threeparttable}
\end{table}
 
The Bayesian inference method presented in the above section is generally valid. When we assume a specific parameterized population model $\mathcal{R}(z)$ as formulated in equations (\ref{merger_rate}-\ref{prob_merger}), the hyperparameters are $\vec{\lambda}=(\mathcal{R}_n,\tau)$. We assume the priors for \(\mathcal{R}_n\), \(\tau\), and \(\eta\) are uniformly distributed in logarithmic space: from 1 to 1000 \(\rm Gpc^{-3}\,yr^{-1}\) for \(\mathcal{R}_n\), from 0.12 to 10 \(\rm Gyrs\) for \(\tau\), and from 0.01 to 10 for \(\eta\).

Table \ref{tab:2_table} gives the boundaries and injected values of these parameters when simulating the catalog. We utilize the {\texttt emcee} algorithm to sample the posterior probability distribution, with a Markov chain length of 1000 \citep[]{2013PASP..125..306F}. We discard the initial 300 samples to ensure that the chain used has reached convergence.
We find that using GW data alone cannot constrain $\tau$ but can provide some constraints on $\mathcal{R}_{n}$; using only sGRB data can constrain the production of $\mathcal{R}_{n}$ and $\eta$, and $\tau$ to some extent, but there exists a strong degeneracy.

\begin{figure*}
    \centering
    \begin{minipage}{0.49\textwidth}
        \centering
        {
            \includegraphics[width=1\linewidth]{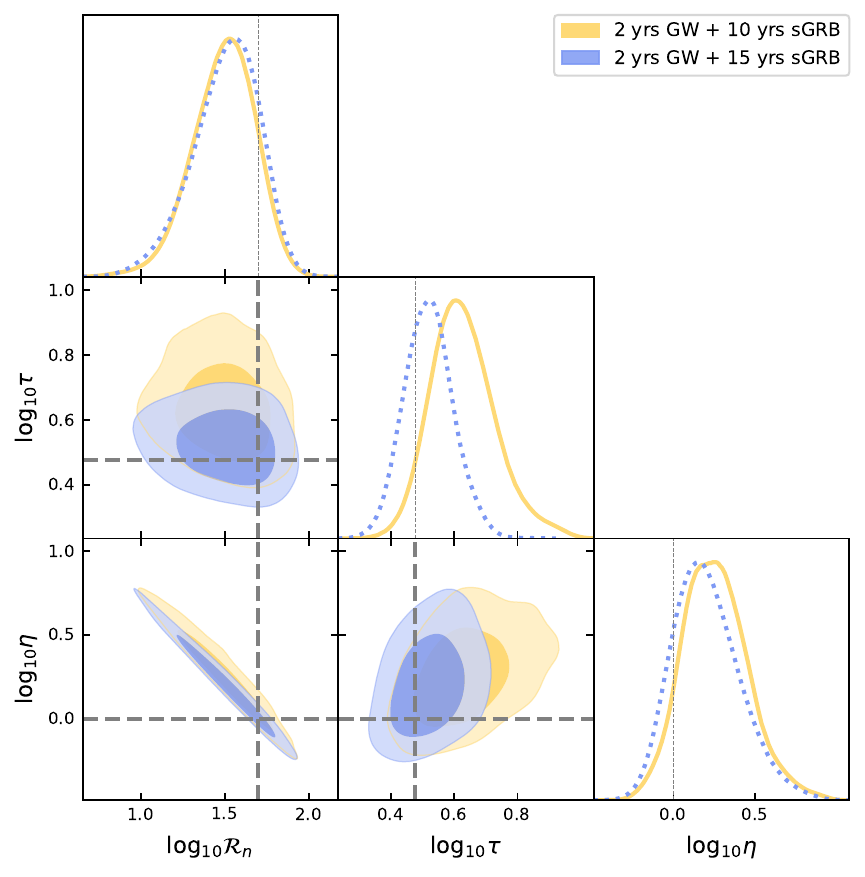}
        }
    \end{minipage}
    \hfill
    \begin{minipage}{0.49\textwidth}
        \centering
        {
            \includegraphics[width=1\linewidth]{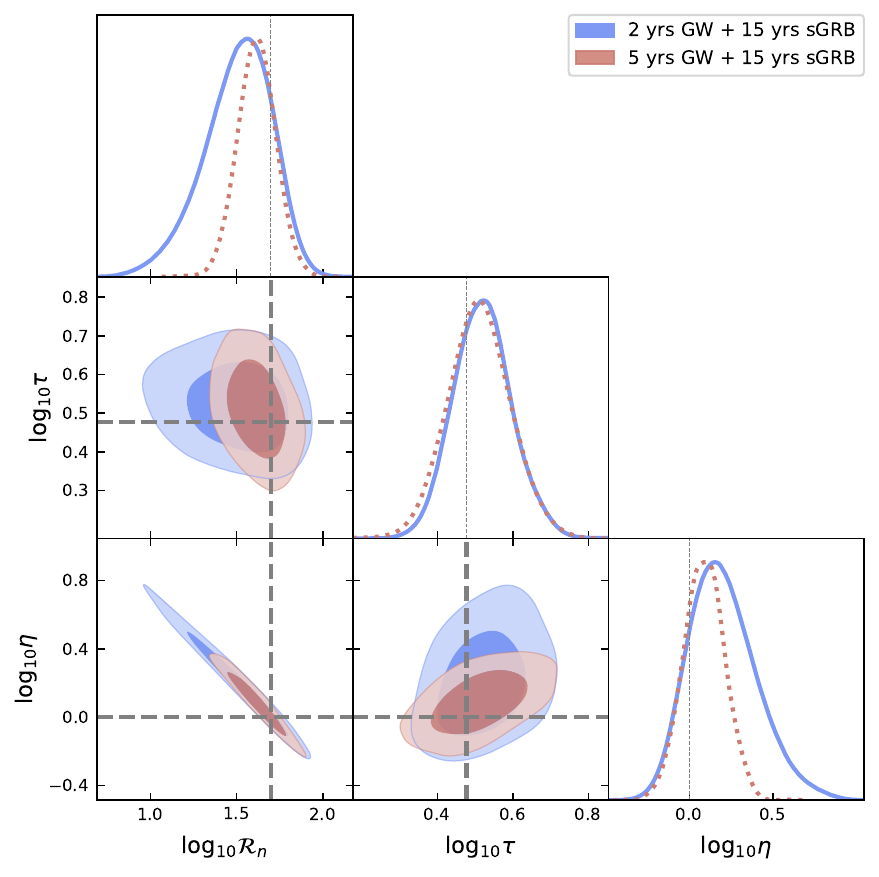}
        }
    \end{minipage}
    
    \vspace{1cm} 
    
    \begin{minipage}{0.49\textwidth}
        \centering
        {
            \includegraphics[width=1\linewidth]{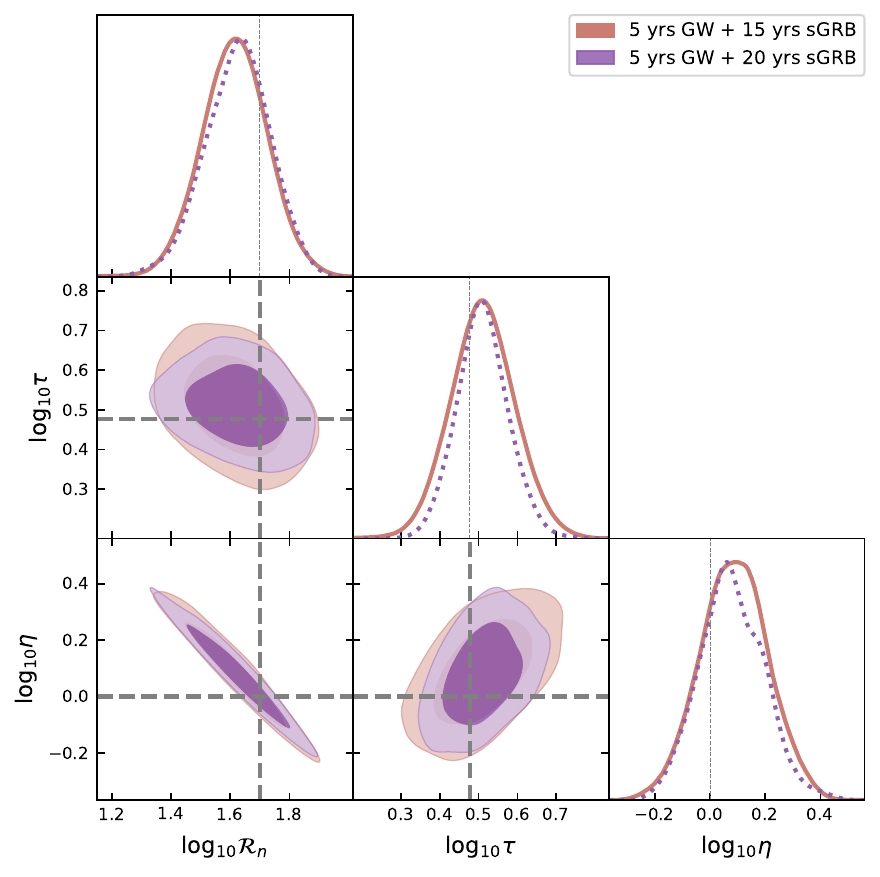}
        }
    \end{minipage}
    \hfill
    \begin{minipage}{0.49\textwidth}
        \centering
        {
            \includegraphics[width=1\linewidth]{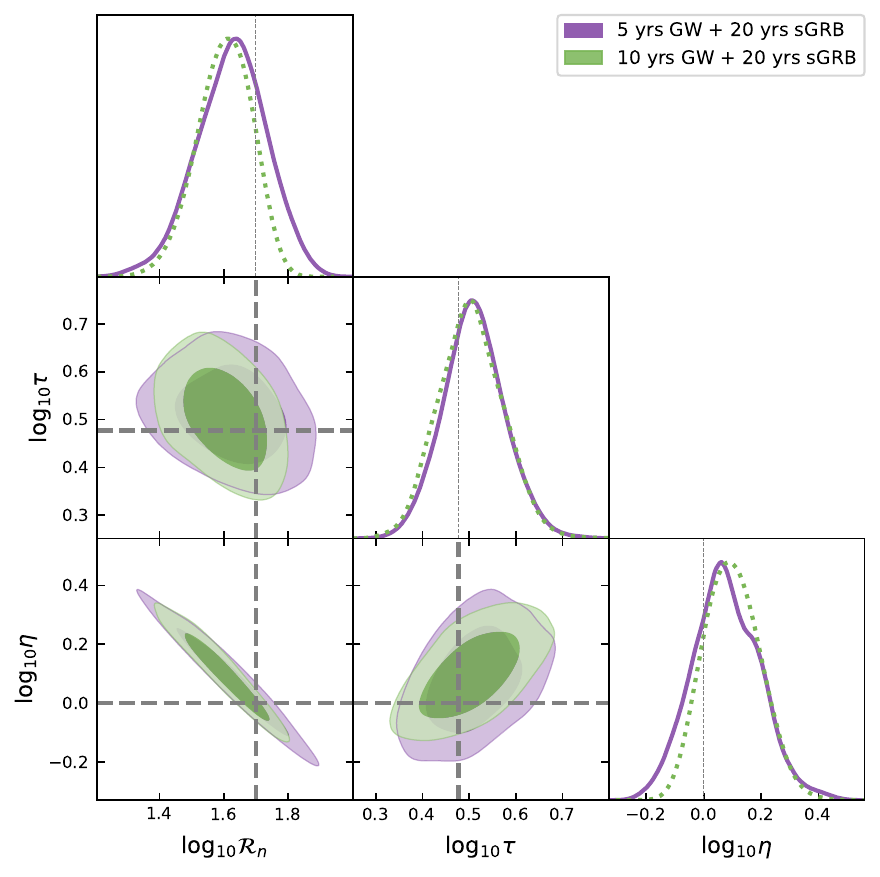}
        }
    \end{minipage}
    
    \caption{The posterior distributions of ${\rm log_{10}}\mathcal{R}_{n}$, ${\rm log_{10}}\tau$, and ${\rm log_{10}}\eta$. The grey dashed lines represent the injected values of the parameters used in the simulation. \textbf{Upper Left:} The simulation data consists of 2 years of GW data and 10 years of sGRB data or 2 years of GW data and 15 years of sGRB data. The dotted line represents the 2 years of GW data and 15 years of sGRB data.  \textbf{Upper Right:} The simulation data consists of 2 years of GW data and 15 years of sGRB data or 5 years of GW data and 15 years of sGRB data. The dotted line represents the 5 years of GW data and 15 years of sGRB data. \textbf{Lower Left:} The simulation data consists of 5 years of GW data and 15 years of sGRB data or 5 years of GW data and 20 years of sGRB data. The dotted line represents the 5 years of GW data and 20 years of sGRB data. \textbf{Lower Right:} The simulation data consists of 5 years of GW data and 20 years of sGRB data or 10 years of GW data and 20 years of sGRB data. The dotted line represents the 10 years of GW data and 20 years of sGRB data.}
    \label{fig:post}
\end{figure*}

Given the operational history of {\it Fermi}/GBM spanning over a decade and the forthcoming attainment of aLIGO's detection sensitivity, we opt to model the joint constraints on the BNS merger rate density evolution using combinations of 2 years of GW data with 10 years of sGRB data, 2 years of GW data with 15 years of sGRB data, 5 years of GW data with 15 years of sGRB data, 5 years of GW data with 20 years of sGRB data, and 10 years of GW data with 20 years of sGRB data. Table \ref{tab:1_table} presents the posterior distributions of the parameters for the various observation durations. Here, $T_{\rm GW}$ and $T_{\rm sGRB}$ denote the observation periods for GW and sGRB, respectively, while $N_{\rm GW}$ and $N_{\rm sGRB}$ represent the number of samples in the catalog. From table \ref{tab:1_table} and Figure \ref{fig:post}, we observe that increasing the number of GW observations leads to a reduction in the relative error of $\mathcal{R}_{n}$, while the relative error of $\tau$ remains relatively stable. Increasing the number of sGRB observations results in little change in the relative error of $\mathcal{R}_{n}$, but a decrease in the relative error of $\tau$. Additionally, the increase in both types of observations leads to a reduction in the relative error of $\eta$. \(\mathcal{R}_{n}\), \(\tau\), and \(\eta\) can be measured with accuracies of 50\%, 18\%, and 45\%, respectively, using 2 years of GW data combined with 15 years of sGRB data. With 5 years of GW data and 20 years of sGRB data, these accuracies improve to 26\%, 16\%, and 28\%, respectively. 

\begin{figure*}
    \centering
    \begin{minipage}{0.49\textwidth}
        \centering
        \includegraphics[width=\linewidth]{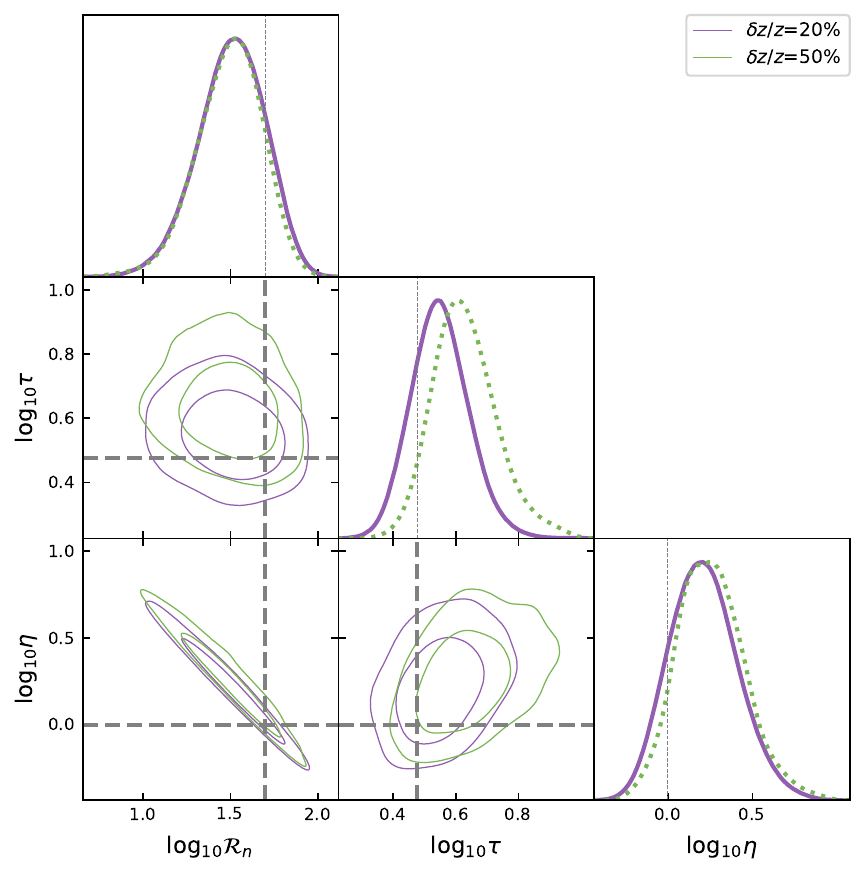}
    \end{minipage}
    \hfill
    \begin{minipage}{0.49\textwidth}
        \centering
        \includegraphics[width=\linewidth]{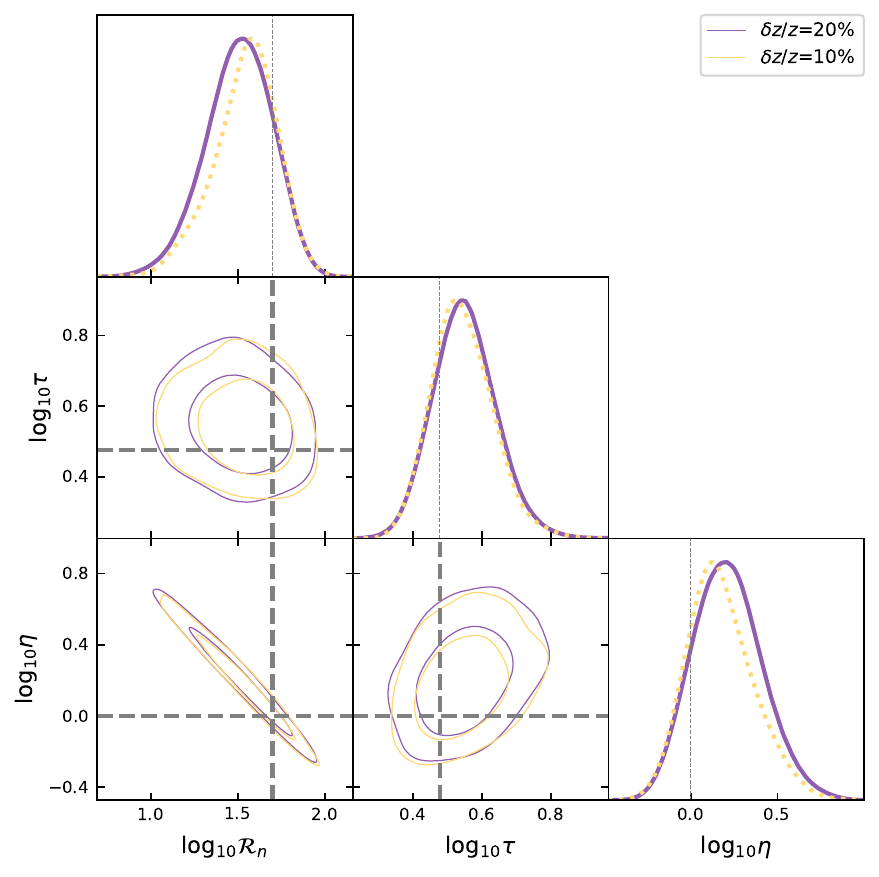}
    \end{minipage}
    \caption{The posterior distributions of ${\rm log_{10}}\mathcal{R}_{n}$, ${\rm log_{10}}\tau$, and ${\rm log_{10}}\eta$. The simulation data consists of 2 years of GW data and 10 years of sGRB data. \textbf{Left:} The estimated redshift's relative error of sGRB we used is 20\% or 50\%. The dashed line represents the 50\% relative error. \textbf{Right:} The estimated redshift's relative error of sGRB we used is 20\% or 10\%. The dashed line represents the 10\% relative error.}
    \label{fig:error}
\end{figure*}

From the upper left panel in Figure \ref{fig:post}, We use 10 years of sGRB data and find significant biases in the constraints of \(\tau\), as well as poorer constraints on \(\eta\) compared to using 15 years of data. Subsequently, we attempt to reduce the redshift's relative error to 20\% or 10\% and observe the disappearance of the previous bias in \(\tau\), as illustrated in Figure \ref{fig:error}. This also indicates that larger redshift errors may lead to biases in the constraints of model parameters, which can be corrected by using more data. Simultaneously, once the biases are eliminated, further reducing the redshift uncertainty will result in a slight improvement in the constraints on the model parameters.

\begin{figure}
    \centering
    \includegraphics[width=.48\textwidth]{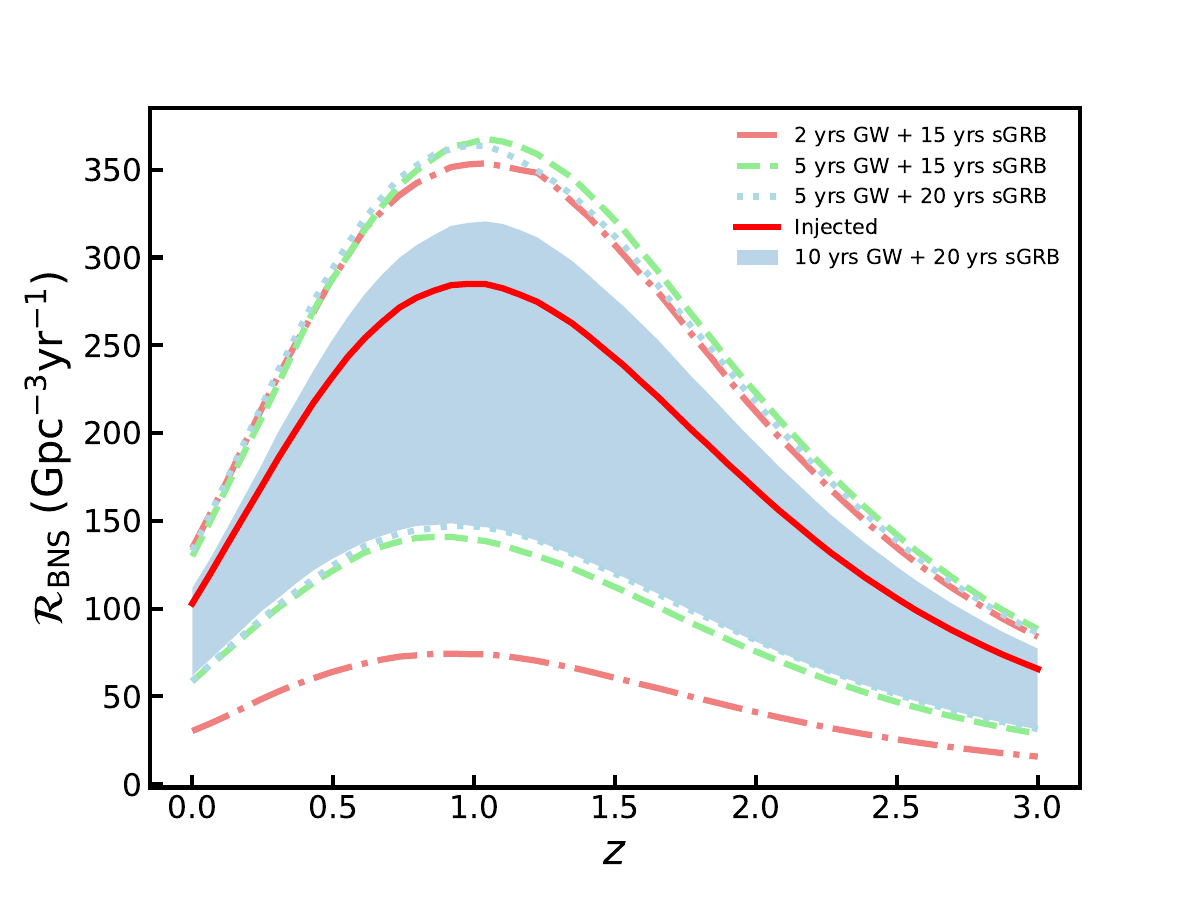}
    \caption{The BNS merger rate density as the function of redshift for different observation durations. The solid red line represents the injected curve based on the parameters used in the simulation. The dashed red line represents the curve obtained using 2 years of GW observations and 15 years of GRB observations. The double-dashed green line represents the curve obtained using 5 years of GW observations and 15 years of GRB observations. The dashed blue line represents the curve obtained using 5 years of GW observations and 20 years of GRB observations. The blue region represents the curve obtained using 10 years of GW observations and 20 years of GRB observations. The region between these lines represents a 90\% confidence level.}
    \label{fig:merger_rate_curve}
\end{figure}

Figure \ref{fig:post} presents the posterior distributions of \({\rm log_{10}}\mathcal{R}_{n}\), \({\rm log_{10}}\tau\), and \({\rm log_{10}}\eta\) based on different GW data and sGRB data. We observe a significant negative correlation between \({\rm log_{10}}\mathcal{R}_{n}\) and \({\rm log_{10}}\eta\), as well as a modest positive correlation between \({\rm log_{10}}\tau\) and \({\rm log_{10}}\eta\). Figure \ref{fig:merger_rate_curve} illustrates the merger rate history for different observation durations. By construction, our model yields a merger‐rate curve of fixed shape, capable only of varying in width (via $\tau$) and normalization (via $R_n$), and cannot reproduce alternative histories (e.g.\ different slope breaks). From this figure, we can see that increasing the amount of GW data narrows the range of the merger rate density within the redshift range of 0-3, especially at lower redshifts between 0-0.2. For instance, extending the analysis from 2 years of GW data to 5 years of GW data. Similarly, using more sGRB data constrains the merger rate density to a smaller range within the redshift range of 0-3, particularly at higher redshifts between 0.5-3, for example, extending the analysis from 15 years of sGRB data to 20 years of sGRB data.
With an increase in the number of sGRBs to a certain threshold, for example, from using 15 years of sGRB data to using 20 years of sGRB data, there is little improvement in the constraint on \(\eta\) and only a minor enhancement in the overall merger rate density evolution with redshift. 
Additionally, we find that with 8 GWs and 571 sGRBs, the merger rate density can be measured to within approximately 50\% accuracy at \(z=0\), and around 50\% accuracy at \(z=1\). With 21 GWs and 761 sGRBs, these accuracies improve to about 35\% at \(z=0\) and 40\% at \(z=1\).

\begin{table*}
	\centering
	\caption{Posterior of the parameters}
	\label{tab:1_table}
	\begin{threeparttable}[b]
	\begin{tabular}{ccccccc} 
		\hline
		$T_{\rm GW}$ (yr) & $T_{\rm sGRB}$ (yr) & $N_{\rm GW}$ & $N_{\rm sGRB}$ & $\mathcal{R}_{n}$ (${\rm Gpc^{-3}\,yr^{-1}}$) & $\tau$ (${\rm Gyr}$) & $\eta$ \\ 
		\hline
		2 & 10 & 8 & 380 & $30.90^{+18.08}_{-9.52}$ & $4.256^{+0.932}_{-0.952}$ & $1.820^{+0.872}_{-0.652}$ \\
		2 & 15 & 8 & 571 & $32.36^{+22.59}_{-9.97}$ & $3.311^{+0.597}_{-0.563}$ & $1.585^{+0.759}_{-0.652}$ \\
		5 & 15 & 21 & 571 & $42.66^{+12.29}_{-9.55}$ & $3.214^{+0.659}_{-0.547}$ & $1.175^{+0.374}_{-0.284}$\\
		5 & 20 & 21 & 761 & $41.69^{+12.01}_{-9.33}$ & $3.236^{+0.540}_{-0.463}$ & $1.202^{+0.383}_{-0.290}$\\
		10 & 20 & 42 & 761 & $40.09^{+9.68}_{-6.67}$ & $3.192^{+0.549}_{-0.469}$ & $1.256^{+0.300}_{-0.281}$\\
		\hline
	\end{tabular}
    \begin{tablenotes}
        \footnotesize
	    \item *  The errors of the parameters are determined by the one-sigma interval of the posterior distribution (68\% credible level).
    \end{tablenotes}
    \end{threeparttable}
\end{table*}

\section{Discussion}

In this study, we systematically demonstrate, through simulations, the feasibility of constraining the BNS merger rate density history using data from both GWs and sGRBs. While previous studies, such as \cite{Hayes2023}, have combined these data sets within a Bayesian framework, the novelty of our work is that rather than assuming  the intrinsic GRB event rate to be identical to the BNS merger rate, we explicitly include their ratio as a parameter to be inferred. By combining GW and sGRB data, we can achieve a more accurate constraint on the redshift evolution of the BNS merger rate density compared to using either data type alone. Meanwhile, the joint data can constrain the proportion $\eta$ of sGRBs produced by BNS mergers. We employed Bayesian inference with inhomogeneous Poisson process likelihood to obtain the posterior distribution of parameters describing the BNS merger rate density history. From the posterior parameter distribution, we can derive the redshift evolution of the BNS merger rate density. At least 15 years of sGRB observations (equivalent to 571 sGRBs) are required to adequately constrain \(\tau\). Using 2 years of GW data and 15 years of sGRB data, \(\mathcal{R}_{n}\), \(\tau\), and \(\eta\) can be measured with accuracies of 50\%, 18\%, and 45\%, respectively. 
The merger rate density can be measured to within approximately 50\% through \(z=0\) to \(z=1\).
We observed that increasing the amount of sGRB data results in a slight improvement in the constraints on the merger rate density, whereas incorporating more GW data significantly enhances the constraints. We also found that the size of the relative error of sGRB estimated redshift can lead to a bias in the constraints on \(\tau\), when the number of sGRBs is small (less than 550). Reducing the relative error in redshifts can help mitigate this bias. 

Future observations of BNS merger GWs will be significantly enhanced by third-generation ground-based detectors such as ET. Simulations indicate that ET could detect BNS merger GWs at a rate of approximately 10$^4$ per year, with a maximum redshift of 10 \citep{2010CQGra..27s4002P, Iacovelli_2022}. It is anticipated that a single year of BNS merger GW observations with third-generation detectors will provide constraints on the BNS merger rate density history that surpass the combined constraints from 20 years of GW and sGRB catalogs presented in this work. However, using only GW catalogs can not constrain \(\eta\). Therefore, if studying the association between sGRBs and BNS mergers is of interest, the combined use of GW and sGRB catalogs offers a viable approach.

Based on the simulations in this work, detecting 42 BNS merger GW sources and 761 sGRBs would constrain the relative uncertainty of $\eta$ to approximately 25\%. This implies that if the most probable value of $\eta$ exceeds 2, it would confirm that a significant fraction of sGRBs originate from non-BNS merger processes, such as collapse scenarios or BH-NS mergers. Conversely, if the most probable value of $\eta$ is below 0.5, it would indicate that many BNS mergers do not produce sGRBs. With future third-generation ground-based GW detectors, the precision of $\eta$ constraints is anticipated to be significantly improved, enabling a more accurate determination of which of the two effects responsible for $\eta \neq 1$ dominates.

In our work, we assumed that sGRBs without redshift measurements and those with redshift measurements have the same distribution over redshift. However, in reality, due to selection effects and the incompleteness of the galaxy catalogues, sGRBs located closer are more likely to have measured redshift, leading to some differences in their distributions. 
Pseudo-redshifts obtained using empirical relations or machine learning methods have shown that the redshift distribution of sGRBs with measured redshifts differs from the overall population \citep[]{2018Ap&SS.363..223Z,2024MNRAS.529.2676A,2024ApJS..271...22D,2024arXiv241013985N}. By studying the differences between the pseudo-redshift distribution and the measured redshift distribution of sGRBs, we can better understand the relationship between these two distributions. This relationship can then be used to simulate samples of sGRBs with and without redshift measurements, helping to mitigate selection effects in this study.

The parameterized population model depends on the specific formulation we assume. We also hope to obtain model independent constraints on $\mathcal{R}(z)$. The binned Gaussian process is one of the non-parametric Bayesian methods, which has the advantage of not requiring many assumptions about the model of BNS mergers \citep[]{2023ApJ...957...37R}. It only assumes that the merger rate density history is a piecewise function and specifies the covariance matrix between the piecewise functions. However, the downside is that it significantly increases computational cost. 

It is worth emphasizing that we assume a structured gaussian jet, and the jet structure of GRBs are still under debate. Alternative angular profiles—most notably broken power-law jets or two‐component (double-Gaussian) jets—exhibit more extended wings than a simple Gaussian core, and would therefore enhance the detectability of off-axis events relative to our current model. For instance, power-law jets decline as $L\propto (\theta_v/\theta_c)^{-k}$ at large angles, boosting the observable fraction of sGRBs at moderate viewing angles, whereas double-Gaussian jets combine a narrow, ultra-relativistic core with a broader, lower-luminosity sheath. Because the sheath component of the double-Gaussian decays more slowly than the single-Gaussian core—but faster than the pure power law—the resulting $P_{\rm det}(\theta_v)$ curve exhibits a moderate “shoulder” at $\theta_v>\theta_{c1}$ (where $\theta_{c1}$ is the narrow core width) that is higher than the Gaussian case yet lower than the power-law case. Nonetheless, the final impact of these alternative models should be investigated in future studies.

Currently, we have only detected two cases of BNS merger GWs and approximately 600 sGRBs detected by {\it Fermi}/GBM. Based on this work, it is anticipated that these GWs and sGRBs can provide some good constraints on the merger rate history and $\eta$. In actual detections, the detection probability of GWs or sGRBs may vary over time. For GWs, we can estimate the S/N at the time by calculating the noise using the real data, allowing us to determine the detection probability at that time. However, for sGRBs, we can only estimate the detection probability over the detection period by studying the variation of their detection rate over time. The hierarchical Bayesian model employed by \cite{2023A&A...680A..45S} serves as a valuable reference method, as it allows for the inference of population models directly from sGRB observational data without requiring specific redshift or jet opening angle measurements. We also plan to adopt hierarchical Bayesian methods to study the population model parameters of BNS mergers using real sGRB and GW data, and this work is currently in progress.

By studying the evolution of the BNS merger rate density history, it is possible to reconstruct the evolution of heavy elements in the Universe caused by BNS mergers and compare it with existing observations \citep[]{2014A&A...564A.134M, 2018ApJ...855...99C}. This can help constrain current models of r-process heavy elements generation from BNS mergers.

\section*{Acknowledgements}

This work is supported by the National Key R\&D Program of China (2021YFA0718500) and the Strategic Priority Research Program of the Chinese Academy of Sciences, Grant No. XDB0550300. S. X. Yi acknowledges the support by the Institute of High Energy Physics (Grant No. E32983U8). E.S.Y. acknowledges support from the “Alliance of International
Science Organization (ANSO) Scholarship For Young Talents.” We also acknowledge funding support from the National Natural Science Foundation of China
(NSFC) under grant No. 12333007.

\section*{Data Availability}

All the data generated via simulation and used in the analysis are available upon request. The code used to generate these data are open sources, \href{https://gw-universe.org/}{https://gw-universe.org/}.


\bibliographystyle{mnras}
\bibliography{example} 

\begin{thebibliography}{}
\makeatletter
\relax
\def\mn@urlcharsother{\let\do\@makeother \do\$\do\&\do\#\do\^\do\_\do\%\do\~}
\def\mn@doi{\begingroup\mn@urlcharsother \@ifnextchar [ {\mn@doi@} {\mn@doi@[]}}
\def\mn@doi@[#1]#2{\def\@tempa{#1}\ifx\@tempa\@empty \href {http://dx.doi.org/#2} {doi:#2}\else \href {http://dx.doi.org/#2} {#1}\fi \endgroup}
\def\mn@eprint#1#2{\mn@eprint@#1:#2::\@nil}
\def\mn@eprint@arXiv#1{\href {http://arxiv.org/abs/#1} {{\tt arXiv:#1}}}
\def\mn@eprint@dblp#1{\href {http://dblp.uni-trier.de/rec/bibtex/#1.xml} {dblp:#1}}
\def\mn@eprint@#1:#2:#3:#4\@nil{\def\@tempa {#1}\def\@tempb {#2}\def\@tempc {#3}\ifx \@tempc \@empty \let \@tempc \@tempb \let \@tempb \@tempa \fi \ifx \@tempb \@empty \def\@tempb {arXiv}\fi \@ifundefined {mn@eprint@\@tempb}{\@tempb:\@tempc}{\expandafter \expandafter \csname mn@eprint@\@tempb\endcsname \expandafter{\@tempc}}}

\bibitem[\protect\citeauthoryear{{Abbott} et~al.,}{{Abbott} et~al.}{2017}]{2017ApJ...848L..12A}
{Abbott} B.~P.,  et~al., 2017, \mn@doi [\apjl] {10.3847/2041-8213/aa91c9}, \href {https://ui.adsabs.harvard.edu/abs/2017ApJ...848L..12A} {848, L12}

\bibitem[\protect\citeauthoryear{{Abbott} et~al.,}{{Abbott} et~al.}{2018}]{2018PhRvL.120i1101A}
{Abbott} B.~P.,  et~al., 2018, \mn@doi [\prl] {10.1103/PhysRevLett.120.091101}, \href {https://ui.adsabs.harvard.edu/abs/2018PhRvL.120i1101A} {120, 091101}

\bibitem[\protect\citeauthoryear{{Abbott} et~al.,}{{Abbott} et~al.}{2021}]{2021ApJ...913L...7A}
{Abbott} R.,  et~al., 2021, \mn@doi [\apjl] {10.3847/2041-8213/abe949}, \href {https://ui.adsabs.harvard.edu/abs/2021ApJ...913L...7A} {913, L7}

\bibitem[\protect\citeauthoryear{{Abbott} et~al.,}{{Abbott} et~al.}{2023}]{2023PhRvX..13a1048A}
{Abbott} R.,  et~al., 2023, \mn@doi [Physical Review X] {10.1103/PhysRevX.13.011048}, \href {https://ui.adsabs.harvard.edu/abs/2023PhRvX..13a1048A} {13, 011048}

\bibitem[\protect\citeauthoryear{{Ahumada} et~al.,}{{Ahumada} et~al.}{2021}]{2021NatAs}
{Ahumada} T.,  et~al., 2021, \mn@doi [Nature Astronomy] {10.1038/s41550-021-01428-7}, \href {https://ui.adsabs.harvard.edu/abs/2021NatAs...5..917A} {5, 917}

\bibitem[\protect\citeauthoryear{{Aldowma} \& {Razzaque}}{{Aldowma} \& {Razzaque}}{2024}]{2024MNRAS.529.2676A}
{Aldowma} T.,  {Razzaque} S.,  2024, \mn@doi [\mnras] {10.1093/mnras/stae535}, \href {https://ui.adsabs.harvard.edu/abs/2024MNRAS.529.2676A} {529, 2676}

\bibitem[\protect\citeauthoryear{{Atal}, {Blanco-Pillado}, {Sanglas}  \& {Triantafyllou}}{{Atal} et~al.}{2022}]{2022PhRvD.105l3522A}
{Atal} V.,  {Blanco-Pillado} J.~J.,  {Sanglas} A.,   {Triantafyllou} N.,  2022, \mn@doi [\prd] {10.1103/PhysRevD.105.123522}, \href {https://ui.adsabs.harvard.edu/abs/2022PhRvD.105l3522A} {105, 123522}

\bibitem[\protect\citeauthoryear{{Berger}}{{Berger}}{2014}]{2014ARA&A..52...43B}
{Berger} E.,  2014, \mn@doi [\araa] {10.1146/annurev-astro-081913-035926}, \href {https://ui.adsabs.harvard.edu/abs/2014ARA&A..52...43B} {52, 43}

\bibitem[\protect\citeauthoryear{{Burns}, {Connaughton}, {Zhang}, {Lien}, {Briggs}, {Goldstein}, {Pelassa}  \& {Troja}}{{Burns} et~al.}{2016}]{2016ApJ...818..110B}
{Burns} E.,  {Connaughton} V.,  {Zhang} B.-B.,  {Lien} A.,  {Briggs} M.~S.,  {Goldstein} A.,  {Pelassa} V.,   {Troja} E.,  2016, \mn@doi [\apj] {10.3847/0004-637X/818/2/110}, \href {https://ui.adsabs.harvard.edu/abs/2016ApJ...818..110B} {818, 110}

\bibitem[\protect\citeauthoryear{{Chen}, {Huang}  \& {Huang}}{{Chen} et~al.}{2019}]{2019ApJ...871...97C}
{Chen} Z.-C.,  {Huang} F.,   {Huang} Q.-G.,  2019, \mn@doi [\apj] {10.3847/1538-4357/aaf581}, \href {https://ui.adsabs.harvard.edu/abs/2019ApJ...871...97C} {871, 97}

\bibitem[\protect\citeauthoryear{{Chen}, {Li}, {Chen}, {Hu}  \& {Liang}}{{Chen} et~al.}{2024}]{2024MNRAS.529.1154C}
{Chen} M.-H.,  {Li} L.-X.,  {Chen} Q.-H.,  {Hu} R.-C.,   {Liang} E.-W.,  2024, \mn@doi [\mnras] {10.1093/mnras/stae475}, \href {https://ui.adsabs.harvard.edu/abs/2024MNRAS.529.1154C} {529, 1154}

\bibitem[\protect\citeauthoryear{{C{\^o}t{\'e}} et~al.,}{{C{\^o}t{\'e}} et~al.}{2018}]{2018ApJ...855...99C}
{C{\^o}t{\'e}} B.,  et~al., 2018, \mn@doi [\apj] {10.3847/1538-4357/aaad67}, \href {https://ui.adsabs.harvard.edu/abs/2018ApJ...855...99C} {855, 99}

\bibitem[\protect\citeauthoryear{{Coward} et~al.,}{{Coward} et~al.}{2012}]{2012MNRAS.425.2668C}
{Coward} D.~M.,  et~al., 2012, \mn@doi [\mnras] {10.1111/j.1365-2966.2012.21604.x}, \href {https://ui.adsabs.harvard.edu/abs/2012MNRAS.425.2668C} {425, 2668}

\bibitem[\protect\citeauthoryear{{Cutler} \& {Flanagan}}{{Cutler} \& {Flanagan}}{1994}]{1994PhRvD..49.2658C}
{Cutler} C.,  {Flanagan} {\'E}.~E.,  1994, \mn@doi [\prd] {10.1103/PhysRevD.49.2658}, \href {https://ui.adsabs.harvard.edu/abs/1994PhRvD..49.2658C} {49, 2658}

\bibitem[\protect\citeauthoryear{{Dainotti} et~al.,}{{Dainotti} et~al.}{2024}]{2024ApJS..271...22D}
{Dainotti} M.~G.,  et~al., 2024, \mn@doi [\apjs] {10.3847/1538-4365/ad1aaf}, \href {https://ui.adsabs.harvard.edu/abs/2024ApJS..271...22D} {271, 22}

\bibitem[\protect\citeauthoryear{{Deng}, {Huang}  \& {Xu}}{{Deng} et~al.}{2023}]{2023ApJ...943..126D}
{Deng} C.,  {Huang} Y.-F.,   {Xu} F.,  2023, \mn@doi [\apj] {10.3847/1538-4357/acaefd}, \href {https://ui.adsabs.harvard.edu/abs/2023ApJ...943..126D} {943, 126}

\bibitem[\protect\citeauthoryear{{Du}, {Yorgancioglu}, {Rao}, {Kumar}, {Yi}, {Zhang}  \& {Zhang}}{{Du} et~al.}{2024}]{2024MNRAS.534.2715D}
{Du} Y.~F.,  {Yorgancioglu} E.~S.,  {Rao} J.~H.,  {Kumar} A.,  {Yi} S.~X.,  {Zhang} S.~N.,   {Zhang} S.,  2024, \mn@doi [\mnras] {10.1093/mnras/stae2261}, \href {https://ui.adsabs.harvard.edu/abs/2024MNRAS.534.2715D} {534, 2715}

\bibitem[\protect\citeauthoryear{{Fong}, {Berger}, {Margutti}  \& {Zauderer}}{{Fong} et~al.}{2015}]{2015ApJ...815..102F}
{Fong} W.,  {Berger} E.,  {Margutti} R.,   {Zauderer} B.~A.,  2015, \mn@doi [\apj] {10.1088/0004-637X/815/2/102}, \href {https://ui.adsabs.harvard.edu/abs/2015ApJ...815..102F} {815, 102}

\bibitem[\protect\citeauthoryear{{Foreman-Mackey}, {Hogg}, {Lang}  \& {Goodman}}{{Foreman-Mackey} et~al.}{2013}]{2013PASP..125..306F}
{Foreman-Mackey} D.,  {Hogg} D.~W.,  {Lang} D.,   {Goodman} J.,  2013, \mn@doi [\pasp] {10.1086/670067}, \href {https://ui.adsabs.harvard.edu/abs/2013PASP..125..306F} {125, 306}

\bibitem[\protect\citeauthoryear{{Goldstein} et~al.,}{{Goldstein} et~al.}{2017}]{2017ApJ...848L..14G}
{Goldstein} A.,  et~al., 2017, \mn@doi [\apjl] {10.3847/2041-8213/aa8f41}, \href {https://ui.adsabs.harvard.edu/abs/2017ApJ...848L..14G} {848, L14}

\bibitem[\protect\citeauthoryear{Hayes et~al.}{Hayes et~al.}{2023}]{Hayes2023}
Hayes L.,  et~al., 2023, \mn@doi [The Astrophysical Journal] {10.3847/1538-4357/acb5b7}, 954, 92

\bibitem[\protect\citeauthoryear{{Hendriks}, {Yi}  \& {Nelemans}}{{Hendriks} et~al.}{2023}]{2023A&A...672A..74H}
{Hendriks} K.,  {Yi} S.-X.,   {Nelemans} G.,  2023, \mn@doi [\aap] {10.1051/0004-6361/202244842}, \href {https://ui.adsabs.harvard.edu/abs/2023A&A...672A..74H} {672, A74}

\bibitem[\protect\citeauthoryear{Iacovelli, Mancarella, Foffa  \& Maggiore}{Iacovelli et~al.}{2022}]{Iacovelli_2022}
Iacovelli F.,  Mancarella M.,  Foffa S.,   Maggiore M.,  2022, \mn@doi [\apj] {10.3847/1538-4357/ac9cd4}, 941, 208

\bibitem[\protect\citeauthoryear{{Lan} et~al.,}{{Lan} et~al.}{2023}]{2023ApJ...949L...4L}
{Lan} L.,  et~al., 2023, \mn@doi [\apjl] {10.3847/2041-8213/accf93}, \href {https://ui.adsabs.harvard.edu/abs/2023ApJ...949L...4L} {949, L4}

\bibitem[\protect\citeauthoryear{{Madau} \& {Dickinson}}{{Madau} \& {Dickinson}}{2014}]{2014ARA&A..52..415M}
{Madau} P.,  {Dickinson} M.,  2014, \mn@doi [\araa] {10.1146/annurev-astro-081811-125615}, \href {https://ui.adsabs.harvard.edu/abs/2014ARA&A..52..415M} {52, 415}

\bibitem[\protect\citeauthoryear{{Margutti} \& {Chornock}}{{Margutti} \& {Chornock}}{2021}]{2021ARA&A..59..155M}
{Margutti} R.,  {Chornock} R.,  2021, \mn@doi [\araa] {10.1146/annurev-astro-112420-030742}, \href {https://ui.adsabs.harvard.edu/abs/2021ARA&A..59..155M} {59, 155}

\bibitem[\protect\citeauthoryear{{Mennekens} \& {Vanbeveren}}{{Mennekens} \& {Vanbeveren}}{2014}]{2014A&A...564A.134M}
{Mennekens} N.,  {Vanbeveren} D.,  2014, \mn@doi [\aap] {10.1051/0004-6361/201322198}, \href {https://ui.adsabs.harvard.edu/abs/2014A&A...564A.134M} {564, A134}

\bibitem[\protect\citeauthoryear{{Mooley} et~al.,}{{Mooley} et~al.}{2018}]{2018Natur.561..355M}
{Mooley} K.~P.,  et~al., 2018, \mn@doi [\nat] {10.1038/s41586-018-0486-3}, \href {https://ui.adsabs.harvard.edu/abs/2018Natur.561..355M} {561, 355}

\bibitem[\protect\citeauthoryear{{Narendra} et~al.,}{{Narendra} et~al.}{2024}]{2024arXiv241013985N}
{Narendra} A.,  et~al., 2024, \mn@doi [arXiv e-prints] {10.48550/arXiv.2410.13985}, \href {https://ui.adsabs.harvard.edu/abs/2024arXiv241013985N} {p. arXiv:2410.13985}

\bibitem[\protect\citeauthoryear{{Petrillo}, {Dietz}  \& {Cavagli{\`a}}}{{Petrillo} et~al.}{2013}]{2013ApJ...767..140P}
{Petrillo} C.~E.,  {Dietz} A.,   {Cavagli{\`a}} M.,  2013, \mn@doi [\apj] {10.1088/0004-637X/767/2/140}, \href {https://ui.adsabs.harvard.edu/abs/2013ApJ...767..140P} {767, 140}

\bibitem[\protect\citeauthoryear{{Planck Collaboration} et~al.,}{{Planck Collaboration} et~al.}{2020}]{2020A&A...641A...1P}
{Planck Collaboration} et~al., 2020, \mn@doi [\aap] {10.1051/0004-6361/201833880}, \href {https://ui.adsabs.harvard.edu/abs/2020A&A...641A...1P} {641, A1}

\bibitem[\protect\citeauthoryear{{Punturo} et~al.,}{{Punturo} et~al.}{2010}]{2010CQGra..27s4002P}
{Punturo} M.,  et~al., 2010, \mn@doi [Classical and Quantum Gravity] {10.1088/0264-9381/27/19/194002}, \href {https://ui.adsabs.harvard.edu/abs/2010CQGra..27s4002P} {27, 194002}

\bibitem[\protect\citeauthoryear{{Ray}, {Hernandez}, {Mohite}, {Creighton}  \& {Kapadia}}{{Ray} et~al.}{2023}]{2023ApJ...957...37R}
{Ray} A.,  {Hernandez} I.~M.,  {Mohite} S.,  {Creighton} J.,   {Kapadia} S.,  2023, \mn@doi [\apj] {10.3847/1538-4357/acf452}, \href {https://ui.adsabs.harvard.edu/abs/2023ApJ...957...37R} {957, 37}

\bibitem[\protect\citeauthoryear{{Rouco Escorial} et~al.,}{{Rouco Escorial} et~al.}{2023}]{2023ApJ...959...13R}
{Rouco Escorial} A.,  et~al., 2023, \mn@doi [\apj] {10.3847/1538-4357/acf830}, \href {https://ui.adsabs.harvard.edu/abs/2023ApJ...959...13R} {959, 13}

\bibitem[\protect\citeauthoryear{{Safarzadeh}, {Berger}, {Ng}, {Chen}, {Vitale}, {Whittle}  \& {Scannapieco}}{{Safarzadeh} et~al.}{2019}]{2019ApJ...878L..13S}
{Safarzadeh} M.,  {Berger} E.,  {Ng} K. K.~Y.,  {Chen} H.-Y.,  {Vitale} S.,  {Whittle} C.,   {Scannapieco} E.,  2019, \mn@doi [\apjl] {10.3847/2041-8213/ab22be}, \href {https://ui.adsabs.harvard.edu/abs/2019ApJ...878L..13S} {878, L13}

\bibitem[\protect\citeauthoryear{{Safarzadeh}, {Biscoveanu}  \& {Loeb}}{{Safarzadeh} et~al.}{2020}]{2020ApJ...901..137S}
{Safarzadeh} M.,  {Biscoveanu} S.,   {Loeb} A.,  2020, \mn@doi [\apj] {10.3847/1538-4357/abb1af}, \href {https://ui.adsabs.harvard.edu/abs/2020ApJ...901..137S} {901, 137}

\bibitem[\protect\citeauthoryear{{Salafia}, {Ravasio}, {Ghirlanda}  \& {Mandel}}{{Salafia} et~al.}{2023}]{2023A&A...680A..45S}
{Salafia} O.~S.,  {Ravasio} M.~E.,  {Ghirlanda} G.,   {Mandel} I.,  2023, \mn@doi [\aap] {10.1051/0004-6361/202347298}, \href {https://ui.adsabs.harvard.edu/abs/2023A&A...680A..45S} {680, A45}

\bibitem[\protect\citeauthoryear{{Symbalisty} \& {Schramm}}{{Symbalisty} \& {Schramm}}{1982}]{1982ApL....22..143S}
{Symbalisty} E.,  {Schramm} D.~N.,  1982, \aplett, \href {https://ui.adsabs.harvard.edu/abs/1982ApL....22..143S} {22, 143}

\bibitem[\protect\citeauthoryear{{Yi}, {Nelemans}, {Brinkerink}, {Kostrzewa-Rutkowska}, {Timmer}, {Stoppa}, {Rossi}  \& {Portegies Zwart}}{{Yi} et~al.}{2022}]{2022A&A...663A.155Y}
{Yi} S.-X.,  {Nelemans} G.,  {Brinkerink} C.,  {Kostrzewa-Rutkowska} Z.,  {Timmer} S.~T.,  {Stoppa} F.,  {Rossi} E.~M.,   {Portegies Zwart} S.~F.,  2022, \mn@doi [\aap] {10.1051/0004-6361/202141634}, \href {https://ui.adsabs.harvard.edu/abs/2022A&A...663A.155Y} {663, A155}

\bibitem[\protect\citeauthoryear{{Zhang} et~al.,}{{Zhang} et~al.}{2021}]{2021NatAs...5..911Z}
{Zhang} B.~B.,  et~al., 2021, \mn@doi [Nature Astronomy] {10.1038/s41550-021-01395-z}, \href {https://ui.adsabs.harvard.edu/abs/2021NatAs...5..911Z} {5, 911}

\bibitem[\protect\citeauthoryear{{Zitouni}, {Guessoum}, {AlQassimi}  \& {Alaryani}}{{Zitouni} et~al.}{2018}]{2018Ap&SS.363..223Z}
{Zitouni} H.,  {Guessoum} N.,  {AlQassimi} K.~M.,   {Alaryani} O.,  2018, \mn@doi [\apss] {10.1007/s10509-018-3449-0}, \href {https://ui.adsabs.harvard.edu/abs/2018Ap&SS.363..223Z} {363, 223}

\makeatother
\end{thebibliography}








\bsp	
\label{lastpage}
\end{document}